\documentstyle[preprint,aps]{revtex}
\begin{document}
\draft
\title {Cluster Variation Method, Pad\'e Approximants and Critical
Behaviour}
\author {Alessandro Pelizzola}
\address {Dipartimento di Fisica and Unit\'a INFM,
Politecnico di Torino, I-10129 Torino, Italy}
\date{\today}
\maketitle
\begin{abstract}
In the present paper we show how
non--classical, quite accurate, critical exponents can be
extracted in a very simple way from the
Pad\'e analysis of the results obtained by mean field like
approximation schemes, and in particular by the
cluster variation method.
We study the critical behavior of the Ising
model on several lattices
(quadratic, triangular, simple cubic and face
centered cubic) and two problems of surface critical behaviour.
Both unbiased and biased
approximants are used, and results
are in very good agreement with the exact or numerical ones.
\end{abstract}

As is well known, mean field like
approximations are very useful tools
in the investigation of phase transitions, but
they fail completely in predicting critical
exponents of low--dimensional systems,
giving results which are independent
of the model and the dimensionality.
This is due to neglecting long range
correlations, and thus they are completely unreliable if one is
interested in the behavior of a model
near critical or multicritical points.

On the other hand, there are some
mean field like approximations which
describe very accurately the low--
and/or high--temperature behaviour of
statistical mechanical models. In
particular, the Cluster Variation Method
(CVM) introduced by Kikuchi\cite{cvm} and reformulated several
times\cite{cumulants} has been
shown\cite{cvmseries} to reproduce exactly many terms of the
high-- and low--temperature expansions of
thermodynamical quantities like specific heat, magnetization and
susceptibility.

In the present paper we show how
this property of the CVM (and also of
other schemes, as we shall see) can be used to build up a very simple
procedure for determining quite accurately
the critical temperature and
critical exponents of a given model,
which relies on the Pad\'e analysis
of low-- and high--temperature results obtained by the CVM.
The plan of our presentation is as
follows: we first review the main ideas
of the CVM and of Pad\'e approximants, then we describe in detail our
technique and several test applications
and finally we discuss results and
possible generalizations.

The Cluster Variation Method in its
modern formulation\cite{cumulants}
can be seen as a
truncation of a cumulant expansion of the functional to be
minimized in the variational principle of
statistical mechanics. The latter
states that the free energy $F$ of a
model system described by the hamiltonian
${\cal H}$ on a lattice $\Lambda$
can be obtained by minimizing the functional
\begin{equation}
F[\rho_\Lambda] = {\rm Tr}(\rho_\Lambda {\cal H} +
k_B T \rho_\Lambda \ln \rho_\Lambda), \label{functional}
\end{equation}
where $k_B$ and $T$ are as customary
Boltzmann's constant and absolute
temperature, with respect to the
density matrix $\rho_\Lambda$, subject to the
constraint ${\rm Tr}\rho_\Lambda
= 1$. Upon introducing the cluster density
matrices $\rho_\alpha = {\rm Tr}_{\Lambda
\setminus \alpha} \rho_\Lambda$,
where
$\alpha$ is a cluster of $n_\alpha$
sites and the trace is performed over all
variables out of $\alpha$, Eq.\ (\ref{functional})
is approximated by a
restricted variational principle for the functional
\begin{equation}
F[\{\rho_\alpha, \alpha \in P\}] = \sum_{\alpha \in P} {\rm
Tr}(\rho_\alpha h_\alpha) + k_B T
\sum_{\alpha \in P} a_\alpha {\rm Tr}
(\rho_\alpha \ln \rho_\alpha), \label{cumfunct}
\end{equation}
where $P$ is a set of \lq\lq maximal\rq\rq\ clusters and all their
subclusters, $h_\alpha$ is the $n_\alpha$--body
interaction contribution due to
the cluster $\alpha$ (maximal clusters
should be taken large enough to
contain all kind of interactions
present in ${\cal H}$), the coefficients
$a_\alpha$ obey\cite{cumulants}
\begin{equation}
\sum_{\beta \subseteq \alpha \in
P} a_\alpha = 1, \qquad \forall \beta
\in P \label{coeffs}
\end{equation}
and the $\rho_\alpha$'s are subject to the constraints
\begin{equation}
{\rm Tr}\rho_\alpha = 1, \forall \alpha \in P
\qquad {\rm and} \qquad \rho_\alpha =
{\rm Tr}_{\beta \setminus \alpha}
\rho_\beta, \forall \alpha \subset \beta
\in P. \label{compatibility}
\end{equation}
Local minima of $F$ can be obtained
in a simple way by means of an iteration
scheme devised by Kikuchi\cite{nim},
the Natural Iteration Method (NIM)
which, for
maximal clusters up to 8--9 sites
as those we used, does not require large
amounts of CPU times.

It should be clear that an approximation
in this scheme is uniquely defined by
the choice of the maximal clusters
(so if the maximal clusters are all the
elementary cells of a simple cubic lattice, we will speak of the cube
approximation, and so on), and it will not be accurate near
critical points, where the correlation
length of the system becomes larger
than the size of the maximal clusters.
Generally speaking, taking larger
maximal clusters will narrow the region
in which the approximation is not accurate.

Let us now turn to a brief review of some basic facts about Pad\'e
approximants\cite{pade}. An
$[L,M]$ Pad\'e approximant to a function $F(z)$ is defined by
\begin{equation}
[L,M] \equiv {P_L(z) \over Q_M(z)}
= {p_0 + p_1 z + p_2 z^2 + \ldots +
p_L z^L \over 1 + q_1 z + q_2 z^2 + \ldots + q_M z^M}. \label{pade}
\end{equation}
The usefulness of
these approximants in statistical
mechanics derives from the fact that a
function with a power--law singular behaviour
\begin{equation}
F(z) \approx A(z)\left(1 - {z \over
z_c}\right)^{-\lambda}, \qquad z \to
z_c^-, \label{powerlaw}
\end{equation}
with $A(z)$ analytic at $z_c$, will
have a logarithmic derivative of the form
\begin{equation}
D(z) \equiv {d \over dz}\ln F(z) \approx - {\lambda \over z - z_c}
[1 + O(z - z_c)], \qquad z \to z_c^-, \label{dlog}
\end{equation}
and a simple pole like that in Eq.\
(\ref{dlog}) can be represented exactly by
Pad\'e approximants. Clearly, $z_c$
and $-\lambda$ are given respectively by
the pole and the corresponding residue of the Pad\'e approximant.
Furthermore, if the exact value (and this is sometimes the
case) or a very accurate estimate
is available for $z_c$, better estimates
of $\lambda$ can be obtained by approximating the function
\begin{equation}
\lambda^*(z) \equiv (z_c - z)D(z) \approx \lambda + O(z - z_c), \qquad
z \to z_c^-. \label{exponent}
\end{equation}
In this case $\lambda$ is estimated directly as the value of the
approximant at $z_c$.
If no estimate of $z_c$ is given
as an input one speaks of {\it unbiased}
approximants, otherwise the approximants are called {\it biased}.

We can now turn to the description
of our technique. To fix ideas, let us
consider the Ising model on a face centered cubic (fcc) lattice. The
hamiltonian is
\begin{equation}
{{\cal H} \over k_B T} = -J \sum_{\langle
i j \rangle} s_i s_j - h\sum_i
s_i, \label{ising}
\end{equation}
where $J$ and $h$ are the (reduced) interaction strength and magnetic
field, $s_i = \pm 1$ is the $z$--component
of a spin 1/2 operator at the
lattice site $i$ and the first summation
is over nearest neighbours (n.n.). Our
first step consists in choosing two
CVM approximations (i.e. two sets of
maximal clusters) $M_1$ and $M_2$
in such a way that $M_2$ can be regarded
as an improvement with respect to
$M_1$. To be concrete, in the following
$M_1$ will be the octahedron plus
tetrahedron approximation proposed by
Aggarwal and Tanaka\cite{cvmseries},
while $M_2$ will be the newly developed {\it
oriented rombohedron} approximation (to be described in detail in a
separate paper\cite{newcvm}). This approximation is obtained
by selecting as maximal clusters all the primitive
rombohedral unit cells of an fcc lattice with a given
orientation (an fcc lattice can be
decomposed into primitive cells of this
shape in four different ways, corresponding
to different orientations).
Since a rombohedron is made up of an octahedron with two tetrahedra
attached on opposite faces, we can
expect this new approximation to improve
with respect to $M_1$. Indeed, this can be easily verified
by comparing different estimates for the critical
temperature: one obtains $T_c = 1/J_c
\simeq 10.03$ from the CVM tetrahedron
approximation, $10.01$ from $M_1$, $9.97$ from $M_2$ and $9.83$
from high--temperature expansions\cite{modpade}.

The next step is to compare results
from $M_1$ and $M_2$ to determine the
temperature (or, equivalently, interaction
strength) range in which $M_1$
is accurate in some sense. Let us consider first the low--temperature
region, in order to obtain estimates
for the critical temperature and the
critical exponent $\beta$ associated to the vanishing of the order
parameter $m = \langle s_i \rangle$.
We have chosen as a measure of the
accuracy of $M_1$ the quantity $\delta
m(J) = |m_1(J) - m_2(J)|$, where
$m_k(J)$ is the value of the order parameter (in zero field)
as given by approximation
$M_k$, which, not too close to the critical point, should be a good
approximation to the absolute error
of $m_1(J)$ with respect to the exact
(unknown) value (it can be checked
on two dimensional problems that this is
a very reasonable assumption). Then
we define a value $J_{\rm min}$ of the
interaction strength such that $\delta
m(J) < \epsilon$ for $J > J_{\rm
min}$, where $\epsilon$ is a small
positive number, and say that $M_1$ (and
consequently also $M_2$) is
{\it accurate} for $J > J_{\rm min}$.
Of course it would be desirable to
take $\epsilon$ very small, since large values of $\epsilon$
cause poor values of the order parameter
to be treated as accurate, but, on
the other hand,
reducing $\epsilon$ narrows the temperature range on which the Pad\'e
analysis will be made. A good compromise, which we have used
throughout this paper is $\epsilon
= 10^{-5}$, which for $M_1$ and $M_2$
as above yields $J_{\rm min} = 0.14$.

We then determine Pad\'e approximants
for the logarithmic derivative of
the magnetization
\begin{equation}
D(z) = {d \over dz} \ln m(z), \label{dlogm}
\end{equation}
where (as customary for low temperature approaches) $z = e^{-J}$ is a
variable which vanishes at zero temperature
and $m(z)$ is given by approximation
$M_1$ (or $M_2$, if this is not very
time--consuming),
by requiring that, for a given pair of positive integers $L$ and $M$,
\begin{equation}
[L,M](z_n) = D(z_n), \qquad z_n =
\exp(J_{\rm min} + n \Delta J), \qquad
n = 0, 1, 2, \ldots L + M. \label{interp}
\end{equation}
The choice of $\Delta J$ is constrained, since the above linear system
becomes ill--conditioned as the interpolation
points become too close, either in
the $z$ direction ($\Delta J$ small) or
in the $D(z)$ direction ($\Delta
J$ large). In Tab.\ \ref{table1} we report
estimates of $T_c$ and $\beta$ obtained
by unbiased $[L,M]$ approximants
for $\Delta J = 0.03$.
They indicate clearly that $T_c \simeq
9.79$ (a strong improvement with
respect to the $M_2$ estimate) and $\beta \simeq 0.31$, which
are in good agreement with the high--temperature
expansion critical temperature
and with $\beta = 0.3258 \pm 0.0044$ (Monte Carlo simulations for the
simple cubic lattice, ref. \cite{mc})
respectively. Different values of
$\Delta J$ in the neighborhood of $0.03$ give similar results.

Once an estimate for $T_c$ has been
obtained, it can be used to construct
biased approximants for the logarithmic
derivative of the high temperature
susceptibility. In the disordered
phase, we measure the accuracy of CVM
approximations by means of $\delta c(J) = |c_1(J) - c_2(J)|$, where
$c_k(J)$ is the n.n. correlation
function $\langle s_i s_j \rangle$ as
given by approximation $M_k$. Requiring
$\delta c(J) < 10^{-5}$ for $J <
J_{\rm max}$ we obtained $J_{\rm
max} = 0.048$. $[L,M]$ approximants have
then been determined for the function
\begin{equation}
\gamma^*(w) = (w_c - w) {d \over dw}\chi(w), \label{gamma}
\end{equation}
where $w = {\rm tanh}(J)$ and $\chi(w)$
is the uniform zero field magnetic
susceptibility. The $L+M+1$ interpolation
points have been defined by $w_n
= {\rm tanh}(J_{\rm max} - n \Delta J), n =
0, 1, \ldots L+M$. For $\Delta J = 0.002$, all the $[L,M]$
approximants with $2 \le
L \le 5$ and $L-1 \le M \le M+1$,
except the $[2,1]$ one, give $\gamma =
1.26$, a result which is not far from the best
estimate $\gamma = 1.2390 \pm 0.0071$
(Monte Carlo simulations for the
simple cubic lattice, ref. \cite{mc}).
A similar analysis on the specific heat gives some
indication for $\alpha \simeq 0.14$,
but data do not accumulate very well.
It should also be noted that biasing
approximants with the high
temperature expansion estimate for the critical temperature improves
results for the critical exponents:
one obtains $\beta \simeq 0.33$ and
$\gamma \simeq 1.24$.

We have done a similar analysis for
the simple cubic lattice, choosing for
$M_1$ the well--known cube approximation
and for $M_2$ a newly developed
(again to be described
in a separate paper\cite{newcvm}) approximation which uses as
maximal clusters both the elementary
cubes and the \lq\lq stars\rq\rq\
formed by one site surrounded by
its six nearest neighbors, which is a
straightforward generalization of
the approximation used by Finel and de
Fontaine\cite{annni} in their investigation
of the two dimensional ANNNI model.
$J_{\min}$ and $J_{\rm max}$ are $0.28$ and $0.13$ respectively.
Low temperature unbiased approximants indicate clearly $T_c \simeq
4.51$, in very good agreement with $T_c = 4.511424 \pm 0.000053$
(Monte Carlo simulations,
ref. \cite{mc}),
while $\beta$ is between $0.30$ and $0.31$. Biased approximants for
$\beta$, however, are in favor of $\beta \simeq 0.31$.
Results from biased high temperature
approximants for the susceptibility
suggest $\gamma \simeq 1.24$.

In Tab.\ \ref{table3} we report results
for two dimensional lattices. In this case,
due to the reduced dimensionality,
the CVM is expected to be less accurate,
but biasing the approximants with
the exactly known critical temperature
yields indeed very good results. We used the approximation $B_{2N}$
proposed by Kikuchi and Brush\cite{bseries}
for the square lattice and a straightforward
generalization of it for the triangular
lattice, choosing $N = 3$ for $M_1$
and $N = 4$ for $M_2$.

Finally, we want to discuss two test
applications of our method to surface
problems. We have considered a semi--infinite
Ising model with a (100) free
surface and unmodified surface coupling,
which is known\cite{semi} to exhibit a
so--called {\it ordinary} transition,
with the surface layer magnetization
$m_1$ vanishing with an exponent
$\beta_1$ which differs from the bulk
exponent $\beta$ and is estimated
to be $0.78 \pm 0.02$ by Monte Carlo
simulations\cite{mcsemi} and $0.816$ by second order
$\epsilon$--expansion\cite{epsilon}.
For this problem we
have developed another new CVM approximation
(again described in detail in
ref. \cite{newcvm}), which we refer to as $4 \times N$,
in which the semi--infinite system is approximated by a film of
$N$ layers, with the topmost layer
representing the free surface, and the
bottom layer constrained to the bulk,
which in turn is studied in the cube
approximation. In this system the
maximal clusters for the CVM are chosen
as those clusters with $4N$ sites,
formed by a column of $N-1$ elementary
cubes. We used $N = 4$ for $M_1$ and $N = 5$ for $M_2$, obtaining
$J_{\rm min} = 0.30$.
Approximants biased with the previously determined bulk critical
temperature for the simple cubic lattice indicate
$\beta_1 \simeq 0.78$, in perfect
agreement with Monte Carlo simulations.

Our last test application aims to
illustrate that the CVM is not the unique
classical approximation on which
our method can be based. In particular we
have considered an approximation recently proposed by Lipowski and
Suzuki\cite{strips}
for two dimensional systems (referred to as the LS approximation),
which has been shown in ref. \cite{mfrg} to yield the
exact boundary magnetization of the square lattice Ising model. It is
essentially a transfer matrix mean
field approximation, where the boundary
magnetization is calculated by considering
two strips of width $N$ and
$N-1$ with periodic boundary conditions
along the infinite direction and
applying the same effective field $h_{\rm
eff}$ at one side of each strip,
where $h_{\rm eff}$ is determined in
such a way that the boundary magnetizations
on the opposite sides of the
strips be equal. The approximation
is no more exact for more complicated
models, but one can expect it to
be quite accurate. We have applied it to
the three--state Potts model in two
dimensions, described by the hamiltonian
\begin{equation}
{{\cal H} \over k_B T} = - K \sum_{\langle i j \rangle}
\delta_{s_i, s_j}. \label{potts}
\end{equation}
We have taken $N = 4$ for $M_1$
and $N = 5$ for $M_2$ ($K_{\rm min} = 0.90$),
and have estimated the boundary magnetization
exponent $\beta_1$ using approximants biased with the exact critical
temperature. Our results indicate clearly $\beta_1
\simeq 0.55$, in very good agreement with $\beta_1 = 5/9$, obtained
combining Cardy's result $\beta_1 = \nu/(3\nu - 1)$\cite{cardy}
with the conjecture
$\nu = 5/6$, supported by many numerical results\cite{potts}.

We have shown how quite accurate
estimates of critical temperatures and
critical exponents can be obtained by a Pad\'e analysis of high-- or
low--temperature results of mean
field like approximations, and especially of
the Cluster Variation Method. The
method is quite simple (some analytical work
can be needed to construct new CVM
approximations, when necessary) and not
at all time--consuming (all the calculations
reported in this paper took a
few hours of CPU time on a DEC Alpha machine), but
nevertheless it yielded very satisfactory
results in several test applications, also in surface problems. There
are also several possible future
developments, among which the calculation of
critical amplitudes and the use of
more sophisticated approximants like
differential and partial differential
approximants (the latter applying to the
study of multicritical phenomena) are worth mentioning.

\narrowtext
\begin{table}
\caption{$(T_c$,$\beta)$ for the
fcc Ising model (unbiased approximants,
$J_{\rm min} = 0.14$, $\Delta J = 0.03$).
\label{table1}}
\begin{tabular}{cccc}
$L$ & $[L,L-1]$ & $[L,L]$ & $[L,L+1]$ \\
\tableline
4 & (9.75, 0.30) & (9.79, 0.31) & (9.78, 0.30) \\
5 & (9.78, 0.31) & (9.78, 0.31) & (9.79, 0.31) \\
6 & (9.78, 0.31) & (9.79, 0.31) & (9.79, 0.31) \\
7 & (9.78, 0.31) & (9.79, 0.31) & (9.79, 0.31) \\
8 & (9.79, 0.31) & (9.79, 0.31) & (9.79, 0.31) \\
\end{tabular}
\end{table}

\begin{table}
\caption{ Results for two dimensional
lattices (biased approximants, exact
$T_c$).
\label{table3}}
\begin{tabular}{ccccc}
Lattice & $J_{\rm min}$ & $\beta$ & $J_{\rm max}$ & $\gamma$ \\
\tableline
Square & 0.53 & 0.123 & 0.30 & 1.73 \\
Triang. & 0.35 & 0.125 & 0.17 & 1.74 \\
\end{tabular}
\end{table}

\end{document}